\documentclass[twocolumn,prb,aps,floatfix]{revtex4}
\usepackage{graphicx}
\usepackage{bm}
\usepackage{epsf}

\begin{document}

\def\bb{\begin{eqnarray}}
\def\ee{\end{eqnarray}}
\title
{Domain-specific magnetization reversals on
a permalloy square ring array} 
\author{D. R. Lee}
\email[Electronic mail: ]{drlee@aps.anl.gov} 
\author{J. W. Freeland}
\author{G. Srajer} 
\affiliation{Advanced Photon Source, Argonne National Laboratory, Argonne, IL 60439}    
\author{V. Metlushko}
\affiliation{Department of Electrical and Computer Engineering,
University of Illinois at Chicago, Chicago, IL 60607}   
\author{Chun-Yeol You}
\affiliation{Department of Physics, Inha University, Incheon 402-751, Korea}

\date{\today}
\begin{abstract}
We present domain-specific magnetization reversals
extracted from soft x-ray resonant magnetic scattering measurement
on a permalloy square ring array.
The extracted domain-specific hysteresis loops reveal
that the magnetization of the domain parallel to the field
is strongly pinned, while those of other domains rotate continuously.
In comparison with the micromagnetic simulation,
the hysteresis loop on the pinned domain indicates
a possibility of the coexistence of the square rings with
the vortex and onion states.
\end{abstract}
\maketitle

A precise control of magnetization reversal involving
well-defined and reproducible magnetic domain states in nanomagnet
arrays is key to future applications, such as high density
magnetic recording\cite{record_jpd_02}
or magnetoelectronic\cite{spintronics_ieee_03} devices.
As topologically various nanomagnets have been
proposed to achieve this, it becomes difficult to characterize
precisely magnetization reversal involving each domain
at small-length scales with conventional magnetization loop measurements.
Moreover, in large-area arrays typically covering areas of
a few square millimeters, extracting overall domain structures
during reversal from magnetic microscopic images is clearly unreliable.
We have presented recently a simple scheme to extract quantitatively
such domain-specific magnetization reversals for nanomagnet arrays from
soft x-ray resonant magnetic scattering (SXRMS) measurements.\cite{lee_ring}
In this paper we will review briefly this scheme and present the explicit
expressions for magnetic form factors, which were deferred
in Ref. \onlinecite{lee_ring}.                          
We will also discuss the extracted domain-specific
hysteresis loops and compare with the micromagnetic simulations.

The array of 20-nm-thick permalloy square rings was fabricated
by a combination of e-beam lithography and lift-off techniques.
From the best fit of x-ray diffraction intensities,
the array period $a$, the width
of the ring $d_1$, and the half of the inner square size
$d_2$, as depicted in Fig. 1, were estimated to be
1151, 162, and 377 nm, respectively, and subsequently the
gap between rings was 73 nm.
X-ray experiments were performed at the soft x-ray beamline
4-ID-C of the Advanced Photon Source.\cite{sector4}  
Circularly polarized soft x-rays were generated by a novel circularly
polarized undulator and tuned to the Ni L$_3$ absorption edge (853 eV) to
enhance the magnetic sensitivity.
\begin{figure}
\epsfxsize=8cm
\centerline{\epsffile{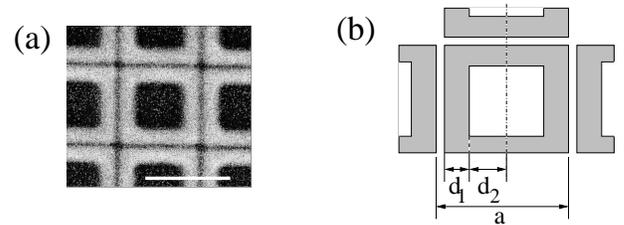}}
\caption{(a) scanning eletron micrograph and 
(b) schematic of the square ring array studied. 
The scale bar is 1 $\mu$m.}  
\end{figure}  

Figures 2(a)-(c) show SXRMS peak intensities measured at various
diffraction orders while field cycling.
All field dependencies show magnetic hysteresis loops
but with different features.
This is due to different magnetic form factors for different
diffraction orders, which reflect nonuniform domain formation
during magnetization reversal, as pointed out
in diffracted MOKE studies.\cite{dmoke_prb_02}
However, as discussed in Ref. \onlinecite{lee_ring},
the difference between the field-dependent intensities of
$I_{np}(H)$ and $I_{pn}(-H)$ should be considered to
take only linear terms of the intensities to the magnetizations
$m_l(H)$ of the $l$-th domain, which are the quantities of interest.
Here $I_{np}$ and $I_{pn}$ represent the intensities measured     
while the field is swept along the negative-to-positive and
positive-to-negative directions, respectively, and
$I_{pn}(-H)$ represents the intensities flipped from
$I_{pn}(H)$ with respect to $H=0$, as shown in Fig. 2(d).
\begin{figure}
\epsfxsize=8cm
\centerline{\epsffile{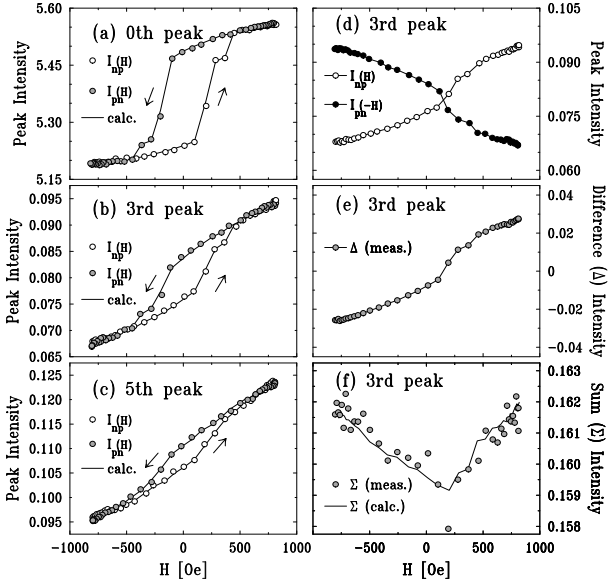}}
\caption{Left panel: SXRMS magnetic hysteresis loops (circles) measured 
at the (a) zeroth, (b) third, and (c) fifth diffraction peaks.  
The solid lines represent the calculated hysteresis loops.
Right panel: (d) SXRMS hysteresis loops with positive-to-negative intensities 
($I_{pn}$) flipped with respect to $H=0$ for the third peak.
Difference (e) and sum (f) intensities between
$I_{np}(H)$ (negative-to-positive) and flipped $I_{pn}(-H)$ intensities
as a function of the applied field.  
The solid lines in (f) represent the calculations.} 
\end{figure}  

These difference intensities $\Delta_{n_x}$ at the $n_x$-order peak, when
normalized by its maximum intensity with a saturation magnetization $m_s$,
can be expressed by a set of linear equations as\cite{lee_ring}
\bb
\frac{\Delta_{n_x}(H)}{|\Delta_{n_x}^{\rm max}|}
= \sum_l B_{n_x l} \frac{m_l(H)}{m_s},
\ee
where
\bb
B_{n_x l}
= \frac{\sum_{|n_y|} F_C \bigl(n_x,|n_y|\bigr)
                F_M^{(l)}\bigl(n_x,|n_y|\bigr) {\cal R}_{n_y}}
            {\sum_{|n_y|} F_C^2 \bigl(n_x,|n_y|\bigr) {\cal R}_{n_y}}.
\ee                         
Applying linear algebra, the normalized magnetizations
$m_{l=1,\cdot\cdot\cdot,N}(H)/m_s$
of $N$ domains can be finally obtained directly from the normalized
difference intensities $\Delta_{n_x}(H)/|\Delta_{n_x}^{\rm max}|$ measured
at $N$ different $n_x$ orders by taking the inverse of
$N\times N$ matrix $B_{n_x l}$.
\begin{figure}
\epsfxsize=8cm
\centerline{\epsffile{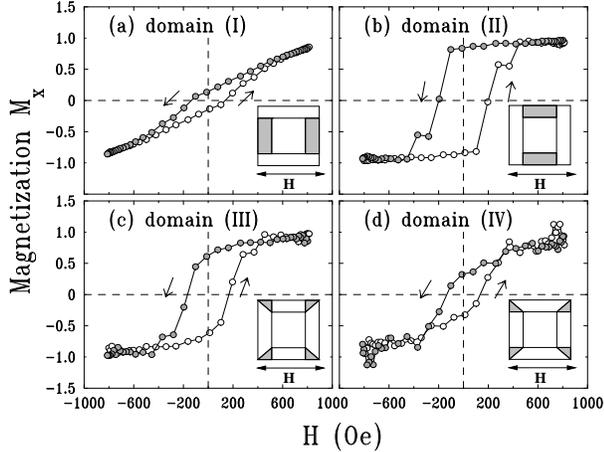}}
\caption{Extracted magnetization reversals along the field direction 
of four types of characteristic domains, which are depicted as gray-filled regions 
in the insets, from SXRMS hysteresis loops at zeroth-,
first-, third-, and fifth-order peaks.  } 
\end{figure}  

For a square ring, the charge form factor $F_C(n_x, |n_y|)$ can be expressed by
\bb
F_C \bigl( n_x, |n_y|\bigr)
&=& {\cal C} \biggl[ {\rm sinc}\Bigl( n_x (\gamma_1 + \gamma_2)\Bigr)
                   {\rm sinc}\Bigl( |n_y| (\gamma_1 + \gamma_2)\Bigr)
        \nonumber \\
        &-& \frac{\gamma_2^2}{(\gamma_1+\gamma_2)^2}
                   {\rm sinc}(n_x \gamma_2)
                {\rm sinc}\Bigl(|n_y|\gamma_2\Bigr) \biggr],
\ee
where ${\cal C}$ is 1 for $|n_y|=0$ and 2 for $|n_y|\neq 0$,
${\rm sinc}(x)= \sin(x)/x$, and $\gamma_{1,2} = 2\pi d_{1,2} /a$.
In our setup, where both incident beam and field directions are parallel to
one of the sides of square rings, there may be four characteristic domains,
as shown in the insets of Fig. 3.    
Each domain consists of two or four subdomains, whose structure-dependent
form factors are identical, and, therefore, its magnetization represents
an average value over subdomains.
The structure-dependent magnetic form factors $F_M^{(l)}$ of these four domains
for $(n_x, |n_y|)$ diffraction order in Eq. (2)
can be explicitly given by
\bb
F_M^{(I)} 
&=& {\cal C} H \bigl(n_x, |n_y|; \gamma_2 \bigr),~~
F_M^{(II)} 
= {\cal C} H \bigl(|n_y|, n_x; \gamma_2 \bigr), \nonumber \\
F_M^{(III)} 
&=& G_-\bigl(n_x,|n_y|\bigr) - G_+\bigl(n_x,|n_y|\bigr) \nonumber \\
            &&~~~~~+ 2 H\bigl(n_x,|n_y|;\gamma_1+\gamma_2\bigr),
            ~{\rm for~} |n_y|\neq 0 \nonumber \\
&=& G_+(0,n_x) - H(0, n_x;\gamma_2), ~{\rm otherwise,} \nonumber \\
F_M^{(IV)} 
&=& G_+\bigl(n_x,|n_y|\bigr) - G_-\bigl(n_x,|n_y|\bigr) \nonumber \\
          &&~~~~~- 2 H\bigl(n_x,|n_y|;\gamma_2\bigr),
            ~{\rm for~} |n_y|\neq 0 \nonumber \\
&=& H(0, n_x;\gamma_1+\gamma_2) - G_+(0,n_x),
            ~{\rm otherwise.}
\ee
where
\bb
H(a,b;c) &=& \frac{c\gamma_1}{(\gamma_1+\gamma_2)^2}
        {\rm sinc}\left( a \frac{\gamma_1}{2} \right) \nonumber \\
&\times&\cos \left( a (\gamma_2 + \frac{\gamma_1}{2})\right)
        {\rm sinc} (b c), \nonumber \\
G_{\pm}(a,b\neq 0) &=&
        \frac{\gamma_1 (\gamma_2+\gamma_1/2)}{(\gamma_1+\gamma_2)^2}
        \left( \frac{a}{b} \pm 1\right)
        {\rm sinc}\left( (a\pm b)\frac{\gamma_1}{2} \right) \nonumber \\
&\times&{\rm sinc}\left( (a\pm b)\left(\gamma_2 + \frac{\gamma_1}{2}\right) \right).
\ee                        

To construct $4\times 4$ matrix $B_{n_x l}$ for four magnetic domains,
zeroth-, first-, third-, and fifth-order peaks were chosen.
Figure 3 shows the extracted magnetization reversals
for each domain using Eqs. (1)-(5).
All magnetizations in Fig. 3 are normalized by the saturation and
represent the components projected along the field direction,
as discussed in Ref. \onlinecite{lee_ring}, of the magnetization vectors.
To confirm these results,
the sum intensities in Fig. 2(f) and the SXRMS hysteresis loops
in Figs. 2(a)-(c) were also generated using the results in Fig. 3.
These calculations show a good agreement with the measurements.
\begin{figure}
\epsfxsize=8cm
\centerline{\epsffile{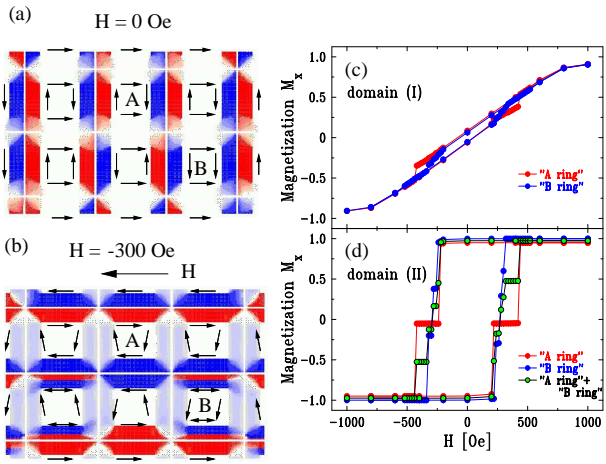}}
\caption{The spin configurations obtained from micromagnetic simulations
with the fields of (a) 0 Oe and (b) -300 Oe.
While the contrast of images varies with the magnitudes of $M_y$
components in (a), that varies with $M_x$ components in (b).
The calculated hysteresis loops using micromagnetic simulations 
in the domains (I) [(c)] and (II) [(d)].}
\end{figure}  

A remarkable difference in the extracted magnetization reversals
is that while the domain (I) perpendicular to the field rotates coherently,
the parallel domain (II) is strongly pinned.
The hysteresis loops calculated from micromagnetic simulations\cite{oommf} 
in Figs. 4(c) and (d) also show clearly this feature.
Interestingly, this is similar to the magnetic domain behaviors
in the antidot arrays,\cite{hole_apl_97,lee_hole}
whose geometry resembles the square
ring array except for narrow gaps between rings.
As discussed in Ref. \onlinecite{lee_hole}, these characteristic domain behaviors
in antidot arrays have been explained by an energetic model including
a shape-induced demagnetization energy contribution.
Though other studies on domain-specific hysteresis loops have not been
reported up to now, this feature is believed to be in common for 
other square ring magnets.\cite{square_ring}

On the other hand, the exptracted hysteresis loop in the domain (II) 
clearly shows plateaus, which
have been observed generally in the vortex state of ring magnets\cite{ring_prl_01} 
and can be also shown in the micromagnetically calculated loop for the ring
with the vortex state (``A'' ring) in Fig. 4(d).
However, the detailed spin configurations from the micromagnetic simulations 
including their $M_y$ components show a noticeable difference from those
in the previous reports on ring magnets.\cite{ring_prl_01}
That is, the vortex- and onion-state rings,
which are marked by ``A'' and ``B'' in Figs. 4(a) and (b), respectively, coexist, 
if we can simply categorize the domain states into only two states 
in terms of whether the magnetizations of the vertical branches of the ring
align parallel or antiparallel to each other 
(at remanence this categorization will be unambiguous).
The overall (average) hysteresis loop in Fig. 4(d) for the domain (II) of both
vortex (A) and onion (B) rings shows then the plateaus with shorter field range
and nonzero magnetization (about a half of the saturation value).
This shows good agreement with the experimentally extracted hysteresis loop
in Fig. 3(b).
We also note that the magnetizations of all adjacent vertical branches 
in Fig. 4(a) align antiparallel due to the dipole interactions between
the rings with very small gaps.

In summary, we presented that domain-specific magnetization reversals can be
extracted directly from SXRMS hysteresis loops measured at various diffraction orders.
From the extracted domain-specific hysteresis loops, we found that
the domain behaviors in a square ring array with very small ring-to-ring gaps
show characteristics of both antidot and ring arrays.
We also found from the comparison with the micromagnetic simulations
that both vortex- and onion-state rings may coexist.

Work at Argonne is supported by the U.S. DOE, Office of Science,
under Contract No. W-31-109-ENG-38.
V.M. is supported by the U.S. NSF, Grant No. ECS-0202780, and    
C.Y. is supported by the National R\&D Project for Nano Science and Technology
of the Ministry of Science and Technology of Korea, M1-0214-00-0001.

\end{document}